# Antiferromagnetism at the $YBa_2Cu_3O_7$ / $La_{2/3}Ca_{1/3}MnO_3$ interface

N. Haberkorn

Universidad Nacional del Sur, Avda. Alem 1253, Bahía Blanca, 8000 Bs. As.,

Argentina.,

Comisión Nacional de Energía Atómica, Centro Atómico Bariloche, S. C. de Bariloche,

8400 R. N., Argentina

J. Guimpel, M. Sirena, and L. B. Steren

Comisión Nacional de Energía Atómica, Centro Atómico Bariloche, S. C. de Bariloche,

8400 R. N., Argentina. and

Instituto Balseiro, Universidad Nacional de Cuyo and Comisión Nacional de Energía

Atómica, S. C. de Bariloche, 8400 R. N., Argentina

W. Saldarriaga, E. Baca and M. E. Gómez

Departamento de Física, Universidad del Valle A. A.,25360 Cali, Colombia.

The magnetic properties of a series of $YBa_2Cu_3O_{7-x}$/$La_{2/3}Ca_{1/3}MnO_3$ (YBCO/$LC_{1/3}$MO) superlattices grown by dc sputtering at high oxygen pressures (3.5 mbar) show the expected ferromagnetic behaviour. However, field cooled hysteresis loops at low temperature show the unexpected existence of exchange bias, effect associated with the existence of ferromagnetic/antiferromagnetic (F/AF) interfaces. The blocking temperature ($T_B$) is found thickness dependent and the exchange bias field ($H_{EB}$) is found inversely proportional to the FM layer thickness, as expected. The presence of an AF material is probably associated to interface disorder and Mn valence shift towards $Mn^{4+}$.

75.70.Cn, 75.30.Vn, 74.76.Bz

The interface between $La_{1-x}A_xMnO_3$ manganites ($LA_xMO$ with $A$: Sr, Ca; $0.2 < x < 0.5$) and $RBa_2Cu_3O_{7-x}$ ($RBCO$ with $R$: Y, Gd) superconductors affects the physical properties of the materials.[1,2] Saturation magnetization ($M_S$), Curie ($T_C$) and superconducting transition ($T_S$) temperatures are reduced from bulk values for $YBCO/LC_{1/3}MO$ superlattices, effect attributed to disorder and oxygen deficiency at the interface.[1-3] In previous work,[4] we studied the structure of $YBCO / LC_{1/3}MO$ superlattices grown on (100) MgO through refinement of the X-ray diffraction patterns. Although the superlattices showed epitaxial growth and lattice parameters similar to those found in films, the interfaces were found not perfect, showing around 1.2 nm roughnesses and 30 % interdiffusion.

In order to study the magnetic structure of the interface, we analysed the magnetic behaviour of the same series of $YBCO/LC_{1/3}MO$ superlattices at low temperatures. In the following they are labelled as $YBCO_N / LC_{1/3}MO_M$, where N indicates the YBCO layer thickness in unit cells (u.c.) (N = 3, 5, 10), and M indicates the $LC_{1/3}MO$ layer thickness in u.c. (M = 9, 15, 30). These values indicate the nominal thickness of the layers, which due to the existence of roughness could fluctuate around N and M in the above specified amount.[4] The total thickness of the superlattices was kept constant at approximately 150 nm by varying the number of bilayers. The superlattices were grown by DC magnetron sputtering in pure oxygen at high pressures (3.5 mbar), as previously reported.[1] Magnetic properties were measured in a commercial SQUID magnetometer, with the applied magnetic field parallel to the superlattice surface. The $T_C$ was estimated from the extrapolation to zero magnetization of the zero field cooled (ZFC) and field cooled (FC) magnetization vs T curves. The $M_S$ was measured with isothermal hysteresis loops at 35 K. The FC hysteresis loops were acquired after cooling in 1T ($H_{FC}$). Transport measurements detect superconductivity below 40 K for $YBCO_5 / LC_{1/3}MO_9$, below 75 K for $YBCO_{10} / LC_{1/3}MO_9$, and below 60 K for $YBCO_{10} / LC_{1/3}MO_{30}$.

Figure 1 shows the ZFC and FC hysteresis loops at 5 K for a $YBCO_3 / LC_{1/3}MO_{15}$ superlattice. The exchange bias signature, i.e. a field shift in the FC loop, is evident and was observed for all studied superlattices. This phenomenon is associated to the existence of a ferromagnetic/antiferromagnetic (F/AF) interface, and is observed when the F layer is cooled through the AF Néel temperature ($T_N$) in the presence of a magnetic field.[5] The unexpected presence of an AF phase can also be inferred from the data in Table 1, which shows $T_C$ and $M_S$. Superlattices $YBCO_5 / LC_{1/3}MO_9$ and $YBCO_{10} / LC_{1/3}MO_9$ are not included since there is a huge uncertainty in $T_C$ and $M_S$, given the smallness of the magnetic signal. The $M_S$ values are strongly reduced from the bulk value ($\approx 580$ emu / $cm^3$) for all superlattices, by a factor between 5 and 10, as previously found for the same system grown on (100) $SrTiO_3$.[1] If we interpret this fact as an amount of FM material transformed into an AF phase, we can define an effective F layer thickness ($t_F$) normalizing $M_s$ to the bulk value. This effective thickness is only 1 u.c. for $YBCO_3 / L_{1/3}CMO_{15}$ and $YBCO_5 / LC_{1/3}MO_{15}$, 5 u.c. for $YBCO_3 / L_{1/3}CMO_{30}$, and 8 u.c. for $YBCO_{10} / LC_{1/3}MO_{30}$. These very low $t_F$ values for the first two superlattices probably imply that the F layers are not continuous in these samples. This possibility is also suggested by the thickness dependence of $H_c$, which reduces with layer thickness, similar to the behaviour in small size particles,[6] and contrary to the $H_c \propto t^{-1}$ one expected for continuous films.[7]

The $T_C$ are also reduced from bulk values for all superlattices, being smaller the thinner the $LC_{1/3}MO$ layers. A reduced $T_C$ for small thickness was also found for $LSr_{0.4}MO$ films,[8] where it was attributed to dimensional effects, structural defects and biaxial strain, and for the $YBCO / LC_{0.67}MO$ system,[3] where it was discussed in terms of stress and proximity effect. The $T_C$ may be influenced simultaneously by stress, valence shift and dimensional effects.

In view of these results we picture the AF phase as mainly distorted manganite due to manganite / cuprate interface disorder. Its origin can be rationalized in several ways. First, it is known that interdiffusion is present. This could originate a new intergrown structured material with an, a priori, unpredictable magnetic behaviour.[4] Second, a shift in the $Mn^{3+}/Mn^{4+}$ relation towards $Mn^{4+}$, and so towards the AF manganite phase, has been found at the $YBCO /LSr_{0.3}MO$ interface,[9] which was related to Ba and La interdiffusion, and which could and probably is present also in the YBCO $/LC_{1/3}MO$ system. Third, a Mn valence shift towards $Mn^{4+}$ can be caused by a change in the crystalline environment, originated either by a change in the oxygen content of the material or by stress at the interfaces. The latter is known to influence the magnetic order in superlattices.[10]

When analysing the behavior of $M_S$ with YBCO layer thickness, we do not find a clear trend. Stadler et al.[9] found a larger displacement towards $Mn^{4+}$ in the $Mn^{3+} / Mn^{4+}$ relation with increasing YBCO thickness in $YBCO/LSr_{0.3}MO$ bilayers, which should result in smaller $M_S$ values. In our results, between $YBCO_3 / LC_{1/3}MO_{15}$ and $YBCO_5 / LC_{1/3}MO_{15}$, $M_S$ is smaller for thicker YBCO layers, but between $YBCO_3 / LC_{1/3}MO_{30}$ and $YBCO_{10} / LC_{1/3}MO_{30}$, $M_S$ is larger for thicker YBCO layers.

Figure 2 shows, for YBCO$_3$/LC$_{1/3}$MO$_{15}$ and YBCO$_3$/LC$_{1/3}$MO$_{30}$, the T dependence of the exchange bias field (H$_{EB}$), H$_{EB}$ = |H$_1$+H$_2$|/2, ZFC hysteresis loop coercive field (H$_{CZ}$) and FC hysteresis loop coercive field (H$_{CF}$), H$_C$ = |H$_1$-H$_2$|/2, where H$_1$ and H$_2$ are the fields for zero magnetization at both branches of the hysteresis loops, measured after saturating the sample with 1T or -1T. For all superlattices, H$_{EB}$ decreases with increasing temperature, reaching zero values at the blocking temperature (T$_B$). We find this temperature independent of YBCO layer thickness and superconducting behavior, and approximately T$_B$ ≈ 30 K for both YBCO$_N$/L$_{1/3}$CMO$_{15}$ superlattices, and T$_B$ ≈ 22 K for both YBCO$_N$/L$_{1/3}$CMO$_{30}$ ones. In the F/AF LC$_x$MO$_N$/LC$_y$MO$_N$ system, it has been reported T$_B$ ≈ 70 K,[11,12] independently of x and y doping values or layer thickness. On the other hand, for LC$_{0.5}$MO/LC$_{1/3}$MO bilayers H. B. Peng et al,[13] showed that T$_B$ and the T dependence of H$_{EB}$ are sensitive to interface roughness. The different T$_B$ for different LC$_{1/3}$MO layer thicknesses again suggests that the AF configuration is different, changing from a continuous layer to a granular system.

For an AF/F/AF system, the exchange energy per unit interface area is given by δE = (M$_S$.t$_F$.H$_{EB}$/2). The H$_{EB}$ values at 5 K, present indeed an inverse thickness, (t$_F$)$^{-1}$, dependence for both LC$_{1/3}$MO thicknesses, with t$_F$ the effective F layer thickness.[5] The estimated δE are independent of the YBCO layer thickness with values ≈ 0.004 erg/cm$^2$ for the YBCO$_N$/LC$_{1/3}$MO$_{15}$ superlattices, and ≈ 0.006 erg/cm$^2$ for the YBCO$_N$/LC$_{1/3}$MO$_{30}$ ones.

The H$_{EB}$ and H$_{CF}$ vs. T curves in Fig. 2 show an exponential dependence instead of the usual power law.[14] This behaviour was previously found in LC$_x$MO/LC$_y$MO superlattices,[12] and attributed to frustration effects resulting from competing F and AF interactions. In our picture, the interfacial disorder is related to YBCO layer roughness, which could locally induce different distortions, and so different local magnetic interactions.

For T lower than $T_B$, $H_{CZ}$ is smaller than $H_{CF}$, which can be attributed to the induction of randomly oriented AF grains on the zero field cooling process. For T higher than $T_B$, where there is no difference between $H_{CZ}$ and $H_{CF}$, a strong domain wall pinning,[15] $H_C \propto (1-T^{2/3})^2$ dependence is found, as in $LSr_{0.4}MO$ films,[7] even for the superlattices with thin $LC_{1/3}MO$ layers. We do not observe a small size particle system dependence,[6] $H_C \propto 1-T^{1/2}$, even though a granular behaviour was expected, which may indicate a domain size distribution.

In conclusion, YBCO/$LC_{1/3}$MO superlattices grown on (100) MgO show an important interface disorder, which could be due to interdiffusion, stress, or oxygen stoichiometry. This disorder induces an AF order at low temperature, which is experimentally signalled by the appearance of exchange bias, with thickness dependent $T_B$ and $H_{EB}$ values. The existence of this AF phase explains the greatly reduced $M_S$ reported values for these superlattices,[1] and suggests that, besides dimensional effects, disorder such as stress and valence distortions must be considered when studying the decrease in $T_C$. It certainly should be considered in the design of devices based on High-Tc / Manganite multilayers.

## Acknowledgments

Work partially supported by ANPCyT grant PICT99-6340, Fundación Balseiro, Fundación Antorchas, Argentina, and COLCIENCIAS, Colombia. NH acknowledges financial support from ANPCyT. JG and LBS are members of CONICET.

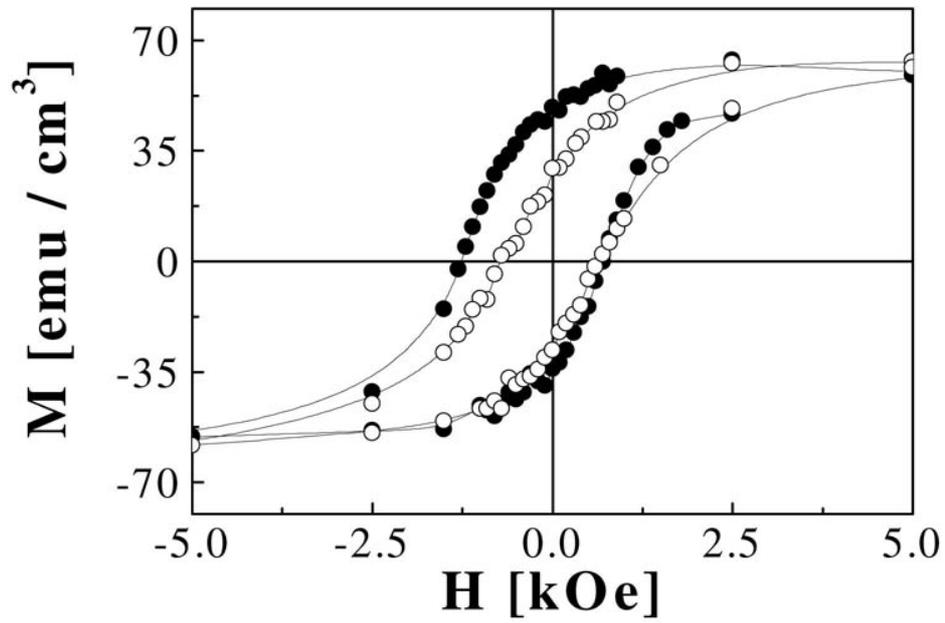

**Figure 1.** Field Cooled, closed circles, and Zero Field Cooled, open circles, hysteresis loops at 5 K for an YBCO$_3$ / LC$_{1/3}$MO$_{15}$ superlattice.

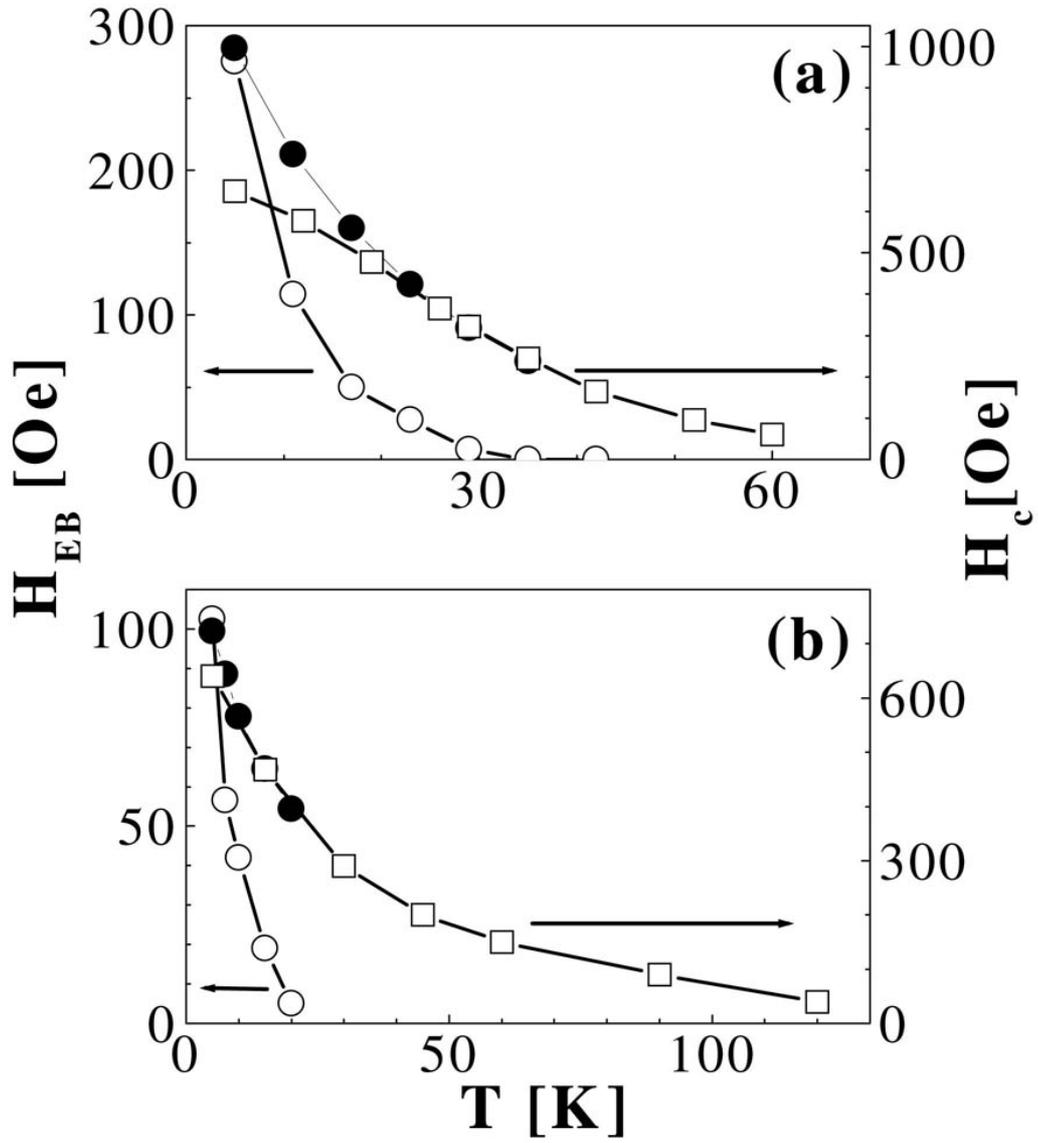

**Figure 2.** Exchange bias field, open circles, field cooled coercive field, solid circles, and zero field cooled coercive field, open squares, versus temperature for (a) YBCO$_3$ / LC$_{1/3}$MO$_{15}$ and (b) YBCO$_3$ / LC$_{1/3}$MO$_{30}$ superlattices.

**Table I.** Curie temperature, $T_C$, and saturation magnetization, $M_S$, for YBCO / $LC_{1/3}MO$ superlattices. $M_S$ was calculated in base to the $LC_{1/3}MO$ volume. The error bars for $M_S$ are estimated from the error in the determination of the effective volume of $LC_{1/3}MO$.

| Superlattice | $T_C$ [K] | $M_S$ [emu / cm$^3$] |
|---|---|---|
| $YBCO_3 / LC_{1/3}MO_{15}$ | 140 ± 15 | 55 ± 10 |
| $YBCO_5 / LC_{1/3}MO_{15}$ | 125 ± 10 | 40 ± 10 |
| $YBCO_3 / LC_{1/3}MO_{30}$ | 220 ± 15 | 100 ± 20 |
| $YBCO_{10} / LC_{1/3}MO_{30}$ | 210 ± 15 | 160 ± 30 |